\documentclass[twocolumn,showpacs,aps,dvips]{revtex4}

\usepackage{graphicx}
\usepackage{epsfig}
\usepackage{amssymb,latexsym,amsmath}

\newcommand{\bs}{\boldsymbol}

\begin{document}

\title{
Relativistic Nucleus-Nucleus Collisions: Zone of Reactions and
Space-Time Structure of a Fireball }

\author{ D. Anchishkin$^1$, A. Muskeyev$^2$, and S. Yezhov$^2$ }

\affiliation{$^1$Bogolyubov Institute for Theoretical Physics,
                 Kiev 03680, Ukraine }
\affiliation{$^2$Taras Shevchenko Kiev National University,
                 Kiev 03022, Ukraine}
\date{\today}

\begin{abstract}
A zone of reactions is determined and then exploited as a tool in
studying the space-time structure of an interacting system formed in
a collision of relativistic nuclei.
The time dependence of the reaction rates integrated over spatial
coordinates is also considered.
Evaluations are made with the help of the microscopic transport
model UrQMD.
The relation of the boundaries of different zones of reactions and
the hypersurfaces of sharp chemical and kinetic freeze-outs is
discussed.
\end{abstract}


\maketitle

{ \bf Introduction.} In the collision of nuclei at
high energies, a strongly excited system of interacting particles is
formed. In fact, a fireball is identified with the zone of reactions
by such a definition, i.e. with a space-time region, in which the
reactions of particles occur. Hence, the zone of reactions must
reflect the space-time characteristics of a fireball, and its study
gives information about the evolution of the interacting system.

While studying the evolution of a fireball, it is important to know
the size of the regions where the majority of the various processes
are running.
Depending on the model describing the system, we can
distinguish the regions of the formation of a fireball, its
isotropization and thermalization, the creation of particles, the
regions of a chemical freeze-out and a kinetic one, etc.
This allows us to conditionally select the stages of evolution of the
system and, hence, to obtain the limits of validity of simple
phenomenological models used for the description of the complicated
physical phenomenon, as well as to describe separate stages of
development of the system in more detail.
In particular, the stages of formation ($\tau\sim 0.1$~fm$/c$) and
thermalization ($\tau\sim 1$~fm$/c$) are most often described with the
use of microscopic
models based on the processes of interaction of quarks and gluons
{\cite{Mueller 2000,Arnold 2004,Rebhan 2004}}.
To describe the stage of spreading of a dense medium
($1\le\tau\le7$~fm$/c$), the relativistic hydrodynamics is most often
in use
{\cite{Russkikh Ivanov 2006,Csorgo Hamma 2003, Sollfrank_et.al 1998}}.
The further evolution of a hadron gas and the process of kinetic
freeze-out ($\tau\sim20$--$25$ fm$/c$) are covered by kinetic models
{\cite{Molnar Gyulassy 2000,Kisiel Florowski 2006}}.
As parameters for the determination of the stages of evolution of a
system, one can take the energy density, mean free path, rate of
collisions of particles, etc.

In the present work, we use the hadron reaction rate (number of
reactions in a unit volume per unit time) in a given
four-dimensional region of space-time  as a parameter of the
spatial evolution of the interacting system.
Such a quantitative estimate allows us to define the reaction zone,
whose study offers the possibility of establishing the space-time
structure of a fireball from the viewpoint of the interaction
intensity at every point of space-time (for a study of the fireball
structure in the momentum space using the number of reactions, see
Refs. \cite{Anch.Yezhov.Musk. 2009p1,Anch.Yezhov.Musk. 2009p2}).
At the same time, the regions of
a fireball can be distinguished by the total number of collisions
which occur there. We use this quantity to determine the boundaries
between different zones of reactions.

\vspace{1.5mm}

{\bf Zones of reactions.} The number of reactions
in the given space-time region can be determined with the use of the
distribution function $f(x,p)$. For example, in the approximation of
two-particle reactions $2\to2$, this function satisfies the
Boltzmann equation \cite{groot}
\begin{equation}
p_1^{\, \mu} \, \partial_\mu \, f_1 \, = \, \int_2\int_3\int_4 \,
W_{12\rightarrow34}\, (f_3\, f_4-f_1\, f_2) \,,
\label{Boltzmann-equation}
\end{equation}
where the right-hand side contains the collision integral.
The quantity $W_{12\rightarrow34}$ is the transition rate which
involves the reaction cross section and the conservation laws,
$\int_i\equiv\int {d^3p_i\over (2\pi)^3E_i}$, $f_i\equiv f(x,p_i)$
are one-particle distribution functions, and
$E_i=\sqrt{m_i^2+{\boldsymbol p}_i^2}$ is the energy of a particle
with momentum ${\boldsymbol p}_i$.
%
\begin{figure*}
\begin{minipage}[l]{.7\textwidth}
\begin{center}
\includegraphics[width=12 cm] {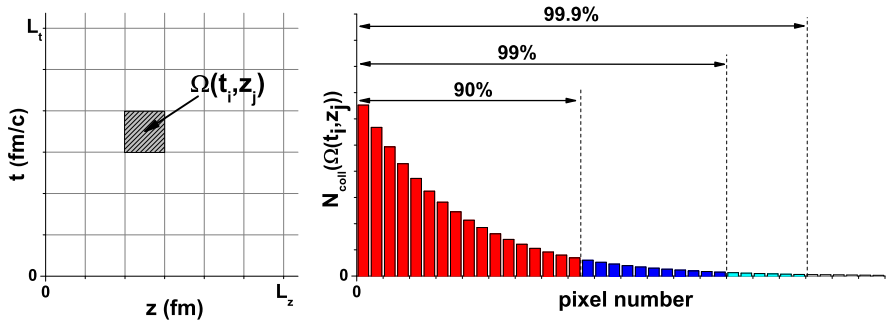}
\end{center}
\end{minipage}
\begin{minipage}[l]{.29\textwidth}
\begin{center}
\caption{(Color online) Algorithm of selection of pixels for the
determination of the zone of reactions. }
\label{fig:ZoR-method}
\end{center}
\end{minipage}
\end{figure*}
%
The probability of a collision of two particles with momenta $\bs
p_1$ and $\bs p_2$ corresponding to the distribution functions $f_1$
and $f_2$, respectively, is determined at a space-time point $x$ as
$\int_3 \int_4 W_{12\rightarrow34}\, f(x,p_1) \, f(x,p_2)$.
By integrating over the momenta of particles, we obtain the rate or
the four-density of reactions at the point $x$:
\begin{equation}
\Gamma(x) = \int_1\int_2 \int_3 \int_4 W_{12\rightarrow34}\,
f(x,p_1) \, f(x,p_2)\,.
\end{equation}
Then, the number of reactions in the given space-time region
$\Omega$ is
\begin{equation}
N_{\rm coll}(\Omega) = \int_\Omega d^4x \, \Gamma(x)
 \, .
\label{coll-number}
\end{equation}
It is seen that the number of reactions in the given space-time
region depends on the four-density of reactions $\Gamma(x)$, which can
be determined in a certain model approximation, e.g., like that in
Refs. {\cite{Eskola Ruuskanen 2007,Tomasik Wiedeman 2003,Hung Shuryak
1998}}.
In particular, $\Gamma(x)$ can be determined with the use of
transport models.

Let us consider a large space-time region containing 99.99 \% of all
two-particle reactions and decays of resonances related to the
event (a particular nucleus-nucleus collision).
That is, the separated space-time region is so large that the
dominant part of all reactions of hadrons occurs in it.
It can be a four-cube of reactions $C_{R}$ with edges $L_i$,
where $i = t,\, x,\, y,\,z$.
To determine the zone of reactions, we divide the cube
into separate equal parts (pixels), i.e., elements of the four-space,
see Fig.~\ref{fig:ZoR-method} (left panel).
Let $\Omega=\Omega(t,{\bs r})$ be the four-volume of a pixel with
coordinates of the center of this volume $(t,{\bs r})$.
Totally, we have $N_{\rm pix} = L_t L_x L_y L_z/\Omega$ pixels.
Then, for each four numbers $(t,{\bs r})$, we can calculate the
absolute number of reactions in the given pixel $\Omega(t,{\bs r})$
by using, e.g., formula (\ref{coll-number}).
After this, we sort the pixels from left to right by the following
hierarchy: from a pair of pixels, the left pixel is that in which a
larger number of reactions has occurred.
The arrangement of pixels is shown in Fig.~\ref{fig:ZoR-method} (right
panel).
The total area of the whole histogram (area covered by all bins) is
equal to the total number of all hadron reactions $N_{\rm tot}$ in
the four-cube of reactions $C_{R}$.

It is obvious that the problem of the determination of a space-time
region, where all collisions have occurred, has a statistical
character. Therefore, it is reasonable to search for this region
with a certain precision. A small part of unaccounted reactions, to
which we can refer also the decays of long-lived resonances, happens
on such time intervals and at such distances which exceed the sizes
of the four-cube of reactions $C_{R}$.

Let us sum the areas of bins beginning from the left according to
the obtained hierarchy. We recall that the area of each bin gives
the number of reactions in the corresponding pixel. In such a way,
we can reach the value of the sum which is equal to a given number
$(\alpha_1\, N_{\rm tot})$, where $\alpha_1$ is a part of the
absolute number of all reactions $N_{\rm tot}$ (see
Fig.~\ref{fig:ZoR-method}).
The hyperspace region which is occupied
by the pixels contributing to this sum gives the reaction zone with
the most intense reaction rate, and an $\alpha_1$ piece of all
hadronic reactions can be attributed to this four-volume.

Exploiting this algorithm, we follow to sum the area of bins to
obtain a next $(\alpha_2\, N_{\rm tot})$ number of reactions.
These
bins determine a neighboring four-volume region, where an $\alpha_2$
piece of all hadronic reactions has occurred. Next, we determine a
four-region which contains an $\alpha_3$ piece of all reactions, and
so on.
Then, $\alpha_1 + \alpha_2 + \alpha_3 + \cdots =~1$.

In the framework of this approach (algorithm), one can analyze a
space-time structure of the fireball for different species of
particles and for different types of reactions, e.g., for elastic or
inelastic reactions, decays, and so on. Obviously, this information
gives insight into the dynamics and the structure of a particular
nucleus-nucleus collision.

\vspace{1.5mm}

{\bf Results of calculations.} To carry out
calculations, we use the transport model UrQMD v2.3 \cite{UrQMD
1998,UrQMD 1999} which allows one to calculate the four-density of
reactions at every point of the space-time region and to select reactions
of a given type and for the given species of particles.
In the present paper, we investigate, first, all possible hadron
reactions and, second, just inelastic hadron reactions.

We take the number of reactions $N_{\rm coll}[\Omega(t,\bs r)]$ in
a pixel $\Omega(t,\bs r)$ as a result of the averaging over 1000
events.
In the calculations, we took a four-cube of reactions $C_{R}$ with the
size of edges $L_i=200$ fm, where $i = t,\, x,\, y,\, z$.

\begin{figure*}
\begin{minipage}[l]{.7\textwidth}
\begin{center}
\includegraphics[width=12 cm] {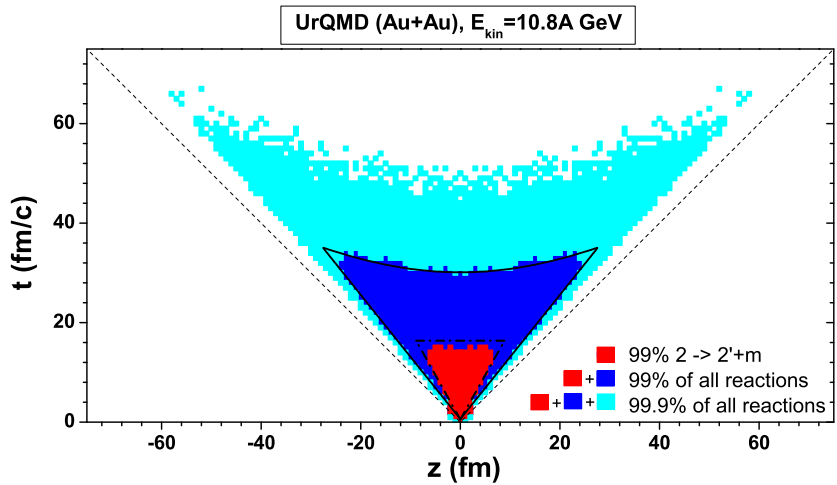}
\end{center}
\end{minipage}
\begin{minipage}[l]{.29\textwidth}
\begin{center}
\caption{(Color online) Projection of the reaction zone on the $z$-$t$
  plane under the AGS (Au+Au at $10.8A$~GeV) conditions.
  The medium-gray (red) region contains 99\% of all inelastic reactions,
  $2\to 2'+m, m\ge 0$.
  The medium-gray (red) and the dark-gray (blue) regions together
  contain 99\% of all hadronic reactions.
  The light-grey (cyan) region contains 0.9\% of all hadron reactions
  only.
  The solid line bounds the region containing 99\% of all reactions.
  The dashed-dotted line bounds the region containing 90\% of
  all reactions.}
\label{fig:ZoR-AGS-ZT-sim}
\end{center}
\end{minipage}
\end{figure*}

\begin{figure*}
\begin{minipage}[l]{.7\textwidth}
\begin{center}
\includegraphics[width=12 cm] {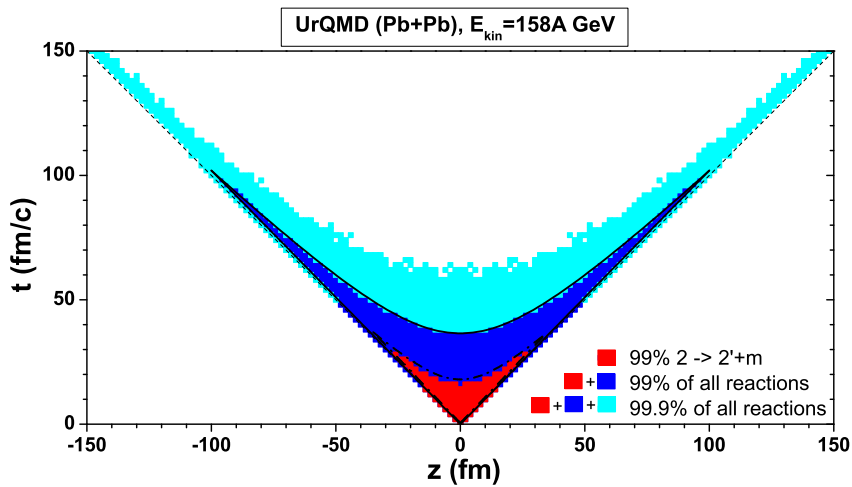}
\end{center}
\end{minipage}
\begin{minipage}[l]{.29\textwidth}
\begin{center}
\caption{(Color online) Same as Fig.~\ref{fig:ZoR-AGS-ZT-sim},
   but for calculations under SPS conditions (Pb+Pb at $158A$~GeV).}
\label{fig:ZoR-SPS-ZT-sim}
\end{center}
\end{minipage}
\end{figure*}

In Figs.~\ref{fig:ZoR-AGS-ZT-sim} and \ref{fig:ZoR-SPS-ZT-sim},
we show the results of calculations for conditions at the
BNL Alternating Gradient Synchrotron (AGS), Au+Au at $10.8A$~GeV,
and at the CERN Super Proton Synchrotron (SPS), Pb+Pb at $158A$~GeV,
in the case of central collisions.
In accordance with the proposed algorithm, we determine the
four-volume which contains 99\% of all hadronic inelastic reactions,
$2\to 2' + m, m \ge 0$ [depicted as the medium-gray (red) area].
We name this zone as a region of {\it hot fireball}.
We determine also a four-volume that contains 99\% of all
possible hadronic reactions which include, of course, the previous
zone.
We name the region of 99\% of all hadronic reactions excluding
the zone of the hot fireball as a {\it cold fireball}
[dark-gray (blue) area].
In fact, one can prefer another precision for the determination of
zones of reactions.

By continuing to move along the hierarchical structure of pixels
from left to right and by gathering a sum of the areas of their
bins, i.e., the numbers of hadronic reactions corresponding to
pixels, we determine those pixels which give, in total, for example,
0.9\% of the total number of all reactions $N_{\rm tot}$.
This additional region can be named as a {\it fireball halo}
[light-gray (cyan) area].
That is, the three space-time regions (hot, cold, halo) cover 99.9\%
of the total number $N_{\rm tot}$ of all hadronic reactions.
%
%

In Figs.~{\ref{fig:ZoR-AGS-ZT-sim}} and {\ref{fig:ZoR-SPS-ZT-sim}},
the rate of reactions is represented in the coordinates $z$-$t$.
That is, we deal with the corresponding projection of the four-volume
of the zone of reactions on the $z$-$t$ plane. To construct
this projection of the three-dimensional spatial pattern onto the
$z$ axis, we sum first all collisions along the transverse
direction at the fixed coordinates $(t,\,z)$, namely [see Eq.
({\ref{coll-number})],
\begin{equation}
\widetilde N_{\rm coll}[\widetilde \Omega(t,z)]
=  \int dx\, dy\, N_{\rm coll}[\Omega(t,x,y,z)]\, .
\label{coll-number-tz}
\end{equation}
Thus, we put a number of reactions $\widetilde N_{\rm coll}$ in
correspondence with the pixel $\widetilde \Omega(t,z)$ with the
coordinates $(t,\, z)$. Then we construct the hierarchy of pixels
$\widetilde \Omega(t,z)$ according to the above-mentioned sorting
algorithm (see Fig.~\ref{fig:ZoR-method}).
It is seen that the interacting system has a comparatively long
lifetime: A hot fireball
decays completely only in the time intervals of the order of $15$~
fm$/c$ for AGS and $35$~fm$/c$ for SPS.
A cold fireball
lives for $30$--$33$~fm$/c$ for AGS and $90$--$100$~fm$/c$ for SPS.
Black solid lines show the approximation of the boundaries of the
reaction zones.

By definition, the zones of hot and cold fireballs together contain
99\% of all reactions.
Moreover, the greater part of reactions outside of these zones
consists of decays of resonances.
If we follow the ``classical'' definition of the sharp kinetic
freeze-out hypersurface as some boundary that separates the
interacting system from the space domain where particles do not
interact, then we can regard the hypersurface which bounds the
region of a cold fireball [the boundary between the dark-grey (blue)
and light-grey (cyan) regions] as the sharp freeze-out hypersurface.

For both collision energies of nuclei under consideration, the upper
parts of freeze-out hypersurfaces in Figs.
{\ref{fig:ZoR-AGS-ZT-sim}} and {\ref{fig:ZoR-SPS-ZT-sim}} are the
space-like hyperbolas, which have the form of constant proper-time
surfaces to within some factor, namely,
$t(z)=A\sqrt{{\tau_0}^2+z^2}$, where $A = 0.65$, $\tau_0 = 46$~fm$/c$
for AGS energies and $A = 0.95$, $\tau_0 = 38$~fm$/c$ for
SPS energies.
From the bottom a cold fireball is  bounded by the
time-like hypersurface, which has the form of a straight line
$t(z)=t_0+{1\over v}z$, where $t_0$ is approximately zero, and
$v=0.8$ for AGS energies and $v=0.98$ for SPS energies.

%
\begin{figure*}
\begin{minipage}[l]{.7\textwidth}
\begin{center}
\includegraphics[width=12 cm] {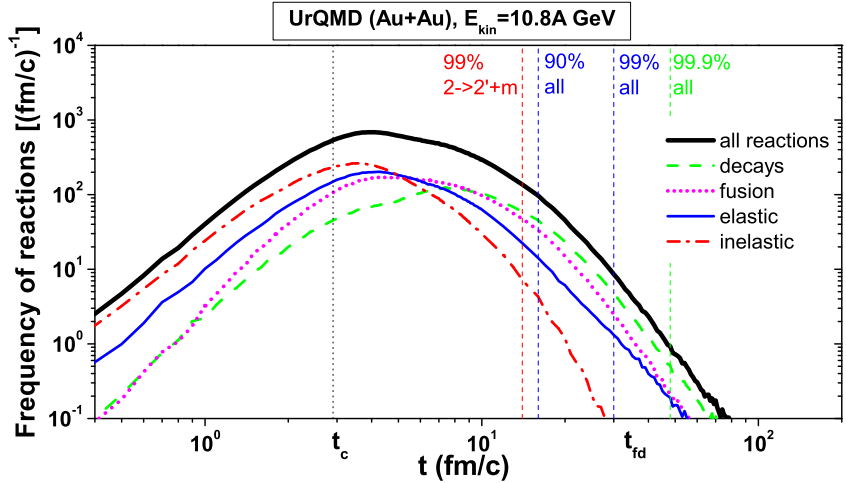}
\end{center}
\end{minipage}
\begin{minipage}[r]{.29\textwidth}
\begin{center}
\caption{(Color online) Frequency of the hadron reactions
(reaction rates integrated over the total spatial volume)
for AGS conditions (Au+Au at $10.8A$~GeV).
  Different curves correspond to different types of reactions.
  Vertical lines indicate the time moments of the margins of zones
  of reactions for the point $z=0$ (see Fig.~\ref{fig:ZoR-AGS-ZT-sim}).}
  \label{fig:ZoR-AGS-T}
\end{center}
\end{minipage}
\end{figure*}
\begin{figure*}
\begin{minipage}[l]{.7\textwidth}
\begin{center}
\includegraphics[width=12 cm] {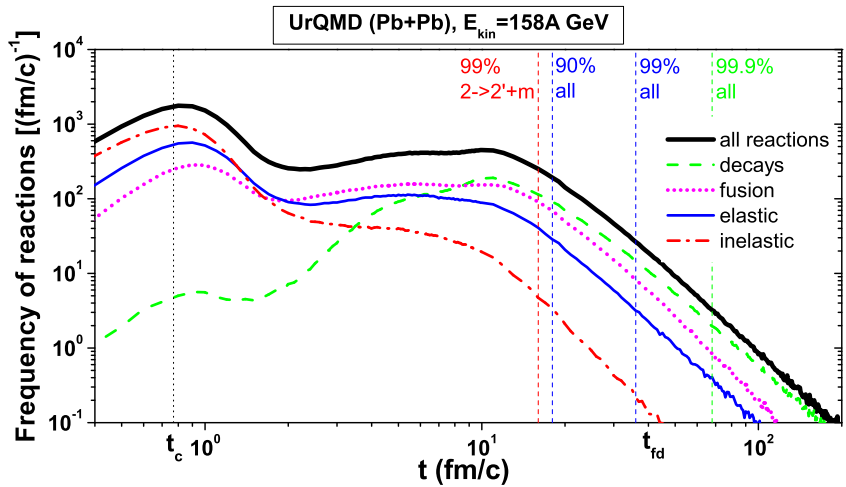}
\end{center}
\end{minipage}
\begin{minipage}[r]{.29\textwidth}
\begin{center}
\caption{(Color online) Same as Fig.~\ref{fig:ZoR-AGS-T}, but for the
SPS conditions (Pb+Pb at $158A$~GeV).}
\label{fig:ZoR-SPS-T}
\end{center}
\end{minipage}
\end{figure*}
%

The results of evaluations of the time dependence of the reaction
rates integrated over the total spatial volume (frequency of
reactions) are depicted in Figs.~\ref{fig:ZoR-AGS-T} and
\ref{fig:ZoR-SPS-T}.
The thick solid line indicates all reaction rates in the fireball, the
thin solid line indicates just the elastic scattering of hadrons
($2\to 2$), the dash-dotted line shows all inelastic reactions ($2
\to 2'+m$, where $m \ge 0$), the dotted line stands for fusion
reactions ($2 \to 1'$), and the dashed line distinguishes decays ($1
\to 2'+m, m \ge 0$). Note that the compound-elastic reactions
(pseudoelastic in Ref.~\cite{Bleicher2002}) such as $2\to \rho \to
2$ are separated in two parts; the first part $2\to \rho$ is
incorporated to the reactions of fusion, and the second part $\rho
\to 2$ to the reactions of decay.

The main feature of the frequency of all reactions (thick solid
lines in Figs.~\ref{fig:ZoR-AGS-T} and \ref{fig:ZoR-SPS-T})
is its increase up to $t\approx 3.9$~fm$/c$ for AGS energies
and $t\approx 0.84$~fm$/c$ for SPS energies.
This can be explained by increasing the number of nucleons as
participants of the reactions, when one nucleus penetrates into
another one.
Indeed, the maximum overlap of two nuclei happens if their centers
coincide.
This time can be estimated as
\begin{equation}
 t_c = \frac{R_{0}}{\gamma}  \frac 1v \,,
\label{tc}
\end{equation}
where $R_0$ is the nucleus radius, $v = p_{0z}/\sqrt{M_N^2 +
p_{0z}^2}$, $\gamma = 1/{\sqrt{1 - v^2}}$, $p_{0z}$ is the initial
nucleon momentum in the c.m. system of two nuclei, and $M_N$ is the
nucleon mass.
In what follows, we name $t_c$ as the {\em fireball formation time}.
For two experiments under consideration, this gives
$t_c = 2.9$~fm$/c$ for AGS ($10.8A$~GeV) and $t_c = 0.77$~fm$/c$ for
SPS ($158A$~GeV).
These values are very close to the time moments which
correspond to the first maximum of the frequency of all reactions
(thick solid lines in Figs.~\ref{fig:ZoR-AGS-T} and \ref{fig:ZoR-SPS-T})
and local maxima of the particular types of reactions for SPS
conditions, see Fig.~\ref{fig:ZoR-SPS-T}.
Slight difference of $t_c$ and the time point of the real maximum
can be explained by some decrease of a nucleon velocity which
is due to inelastic and elastic reactions (stopping) of nucleons.
The same behavior can be seen as well in Ref. {\cite{Bleicher2002}},
Fig.~1, obtained from UrQMD calculations for SPS conditions (Pb-Pb,
$E_{\rm kin} = 158A$~GeV).

Inelastic nucleon collisions dominate at the first stage,
$t \lesssim t_c\,$, of nucleus-nucleus collision.
At later times, the elastic and fusion reactions become more
significant.
We note that for both energy conditions, the decay processes become
the dominant ones after $t\thickapprox 10$~fm/$c$.

After the full overlap of the nuclei, the created system begins to expand
in space, which results in a decrease of the reaction rates.
At the same time, the number of secondary particles still increases,
resulting in an increase of the total reaction integral rate.
Hence, the rate of expansion of the system and its ratio to the
creation rate of secondary particles will determine the result of
the competition of these two tendencies. A similar competition was
investigated in Ref.~\cite{Hung Shuryak 1998}, where on this basis some
conclusions about freeze-out were made.

The number of secondary particles (mainly $\pi$ mesons) is
approximately $\langle n_\pi \rangle \approx 1.6$ per nucleon for
AGS conditions and $\langle n_\pi \rangle \approx 6$ for SPS
conditions.
Such an increase of the number of secondary particles
with the collision energy leads to the sufficient difference of
dependences of the frequency of reactions on time.
Namely, for AGS conditions after $t_c$, the the frequency of all
reactions goes down, which results in {\em fireball division} into two
parts at the time moment $t = t_{\rm fd}$ and the further breakup.
The {\em fireball division time} is defined as the minimum value of
time on the space-like hypersurface, which bounds the region of the
cold fireball (blue area) from above, i.e.,
$t_{\rm fd}\equiv t(z)\big|_{z=0}$
(see Figs. {\ref{fig:ZoR-AGS-ZT-sim}} and {\ref{fig:ZoR-SPS-ZT-sim}}).
For SPS conditions, one can see the second local maximum
of the frequency of reactions (see thick solid line in
Fig.~\ref{fig:ZoR-SPS-ZT-sim}) at
$t_M \approx 10.6$~fm$/c$ which is a consequence of a large number of
reactions with secondary particles.
%
This results in an increase of the {\em fireball lifetime}:
\begin{equation}
\tau = t_{\rm fd} - t_c \,.
\end{equation}
We note that the time moment $t_{\rm fd}$ depends very weakly on the
collision energy.

It is seen that at $z=0$ in the c.m. system after the time moment
$t_{\rm fd}$, the rates of elastic and inelastic reactions vanish.
That is, since this moment, the system behavior is determined mainly
by the individual properties of particles (basically resonances).
That is why, in spite of the sufficient difference of collision
energies of the experiments under consideration, the times $t_{\rm
fd}$ are approximately the same. If we compare the longitudinal
sizes of the fireballs $2 R_z$ at the time moment $t=t_{\rm fd}$, we
see that they are approximately the same and equal to
$R_z = v \, t_{\rm fd}$ [$v$ is defined in Eq.~(\ref{tc}), see
Figs.~\ref{fig:ZoR-AGS-ZT-sim} and \ref{fig:ZoR-SPS-ZT-sim}].
This fact can explain the weak dependence of the pion interferometric
radius $R_{L}$ on the beam energy, $R_{L} \propto R_z$
\cite{Chen2009}.
It can be claimed that the fireball achieves its
maximum longitudinal size at the time moment $t=t_{\rm fd}$, when it
is divided into two parts.

\vspace{1.5mm}

{\bf Discussion and Conclusions.} The main idea of
the method proposed for analyzing the fireball structure is the
determination of the hierarchy of hadronic reactions in accordance
with their intensity (rate), see Fig.{\ref{fig:ZoR-method}}.
In view of the hierarchy obtained, we formulate an algorithm for the
determination of the zones of reactions of the interacting system:
we divide the fireball space-time volume into different regions.
{\it By definition, we recognize a specific zone of reactions
(elastic, inelastic, etc.) as a four-volume, where 99\% of
a given specific type of reactions have occurred. }

In the present microscopic study, we separate a fireball into the
following regions, which characterize its  evolution (see
Figs.~\ref{fig:ZoR-AGS-ZT-sim} and \ref{fig:ZoR-SPS-ZT-sim}):
(1) a hot fireball region, where 99\% of all inelastic hadronic
reactions have occurred (medium-gray or red),
(2) a cold fireball region (dark-gray or blue), which together
with the hot fireball contains 99\% of all hadronic reactions $N_{\rm tot}$,
and (3) a fireball halo, where 0.9\% of all hadronic reactions,
i.e., $0.009 N_{\rm tot}$, have occurred (light-gray or cyan).
Two last regions together are a space-time region containing
the hadron-resonance gas, and the reactions in this region are
mainly presented by decays of resonances.

An important question that can be clarified by the study of the
zone of reactions is how the space-time boundary of a fireball is
related to the so-called sharp freeze-out hypersurface.
In the literature, the sharp freeze-out hypersurface is defined
with the help of some parameter $P(t,\bs r)$ which takes the
critical value $P_c$ on the hypersurface.
That is, the equation of the
hypersurface has form $P(t,\bs r)=P_c$. As such a parameter, one
may choose the energy density $\epsilon(t,\bs r)$ \cite{Russkikh
Ivanov 2006,Sollfrank_et.al 1998}, temperature $T(t,\bs r)$
\cite{mclerran-1986,Huovinen 2007}, density of particles
$n(t,\bs r)$ \cite{CERES}, etc. {\em It is necessary to note that any
definition of the sharp freeze-out is possible just with chosen
accuracy}.

One can follow the ``classical'' definition: the sharp kinetic
freeze-out hypersurface is an imaginary hypersurface, outside of
which there are no collisions between particles of the system. In
other words, the sharp kinetic freeze-out hypersurface is determined
as a surface bounding the space-time region, in which almost all
collisions between hadrons of the system happened (for example,
99\%).
Thus, we can identify the reaction zone boundary and the sharp
kinetic freeze-out hypersurface.
Then the
boundary between the zone of a cold fireball and the fireball halo,
can be interpreted as a sharp kinetic freeze-out hypersurface
(see Figs.~\ref{fig:ZoR-AGS-ZT-sim} and \ref{fig:ZoR-SPS-ZT-sim}).
In the coordinates $(t,\,z)$, the space-time part of this hypersurface
is a hyperbola and has the form $t(z)=A\sqrt{{\tau_0}^2+z^2}$, where
$A = 0.65$, ${\tau_0} = 46$~fm$/c$
for the AGS energy ($E_{\rm kin} = 10.8A$~GeV) and
$A = 0.95$, ${\tau_0} = 38$~fm$/c$
for the SPS energy ($E_{\rm kin} = 158A$~GeV).
It is clear that the {\em fireball division time} is related to the
parameters of the hyperbola in the following way
$t_{\rm fd} = A\, \tau_0$ and we obtain $t_{\rm fd} \approx 30$~fm$/c$
for AGS ($E_{\rm kin} = 10.8A$~GeV) and $t_{\rm fd} \approx 36$~fm$/c$
for SPS ($E_{\rm kin} = 158A$~GeV).

The lower time-like hypersurface bounding a cold fireball has the
form of a straight line $t(z)=t_0+{1\over v}z$, where $t_0$ is
close to zero, and $v = 0.8$ for AGS energies and $v = 0.98$ for
SPS energies.
For AGS energies, $E_{\rm kin} = 10.8A$~GeV, the time-like boundaries
of the reaction zones differ significantly from one another and the
light cone (see Fig. \ref{fig:ZoR-AGS-ZT-sim}).
However, at higher SPS energies, for example, at
$E_{\rm kin} = 158A$~GeV, the time-like hypersurfaces bounding all
three zones of a fireball practically coincide with one another and
are close to the light cone (see Fig.~\ref{fig:ZoR-SPS-ZT-sim}).
Thus, we can predict that this behavior will result in all
time-like hypersurfaces merging on the energies available at the
BNL Relativistic Heavy Ion Collider (RHIC) and coinciding with
the light cone.

On the base of the same ideology, we can define the sharp chemical
freeze-out hypersurface.
Namely, the hypersurface separating the zones of the hot and cold
fireballs can be associated with the sharp chemical freeze-out with an
accuracy of 99\% (we assume that the chemical freeze-out occurs when
the inelastic reactions are completed \cite{Heinz 2001}).

On the other hand, if one deals with a continuous freeze-out, for
instance, by fixing the coordinates of the last collisions, then the
set of these points (with an accuracy of 99\%) will be inside the
boundaries we determined above.

On the basis of the analysis of the time dependence of the frequency
of reactions (see Figs.~\ref{fig:ZoR-AGS-T} and
\ref{fig:ZoR-SPS-T}), we conclude that there are two specific time
points in the evolution of a hadron fireball:
the {\em fireball formation time}
$t_c$ defined as the time of the full overlap of two nuclei
[Eq.~(\ref{tc})] and the {\em fireball division time} $t_{\rm fd}$ which
corresponds to the separation of the fireball into two individual
parts.

\vspace{1.5mm}

{\bf Acknowledgments.}
The authors thank L.~McLerran, P.~Romatschke, and V.~Magas for
useful discussions.
The authors also greatly appreciate the valuable comments made by the
referee.


\end{document}